\documentclass[useAMS,usegraphicx,usenatbib,a4]{mn2e}

\usepackage{times}

\title[Mira variables in the Galactic bulge]
{Mira variables in the Galactic bulge with OGLE-II data}
\author[N. Matsunaga et al.]
{Noriyuki Matsunaga$^{1}$\thanks{E-mail:matsunaga@ioa.s.u-tokyo.ac.jp},
Hinako Fukushi$^{1}$, and Yoshikazu Nakada$^{2,1}$\\
$^{1}$ Institute of Astronomy, School of Science, The University of Tokyo,
Osawa 2-21-1, Mitaka, Tokyo 181-0015, Japan\\
$^{2}$ Kiso Observatory, School of Science, The University of Tokyo,
Mitake, Kiso, Nagano 397-0101, Japan}

\begin{document}

\date{Accepted 2005 August 20. Received 2005 August 10; in original form 2005 May 17.}

\pagerange{\pageref{firstpage}--\pageref{lastpage}} \pubyear{2005}

\maketitle

\label{firstpage}

\begin{abstract}
We have extracted a total of 1968 Mira variables from
the Optical Gravitational Lensing Experiment II (OGLE-II) data base
in the Galactic bulge region. 
Among them, 1960 are associated with 2 Micron All-sky Survey (2MASS) 
sources, and 1541 are further identified with
{\it Midcourse Space Exploration (MSX)} point sources.
Their photometric properties are compared with those of 
Mira variables in the Large and Small Magellanic Clouds.
We have found that mass-losing stars with circumstellar matter are reddened
such that the colour dependence of the absorption coefficient is
similar to that of interstellar matter.
We also discuss the structure of the bulge.
The surface number density of the bulge Mira variables is well correlated
with the 2.2-$\mu$m surface brightness obtained by
the {\it Cosmic Background Explorer (COBE)} satellite.
Using this relation, the total number of Mira variables in the bulge 
is estimated to be about 6 $\times 10^5$.
The $\log P$-$K$ relation
of the Mira variables gives their space distribution which
supports the well-known asymmetry of the bar-like bulge.
\end{abstract}


\section{INTRODUCTION\label{sec:Intro}}

Mira variables are believed to be in the last stage of 
Asymptotic Giant Branch (hereafter AGB) evolution. They have
larger amplitudes (e.g. $> 2.5$~mag in $V$-band) than other red variables,
i.e. semi-regular variables or irregular variables. 
As a result of the short lifetimes of Mira variables, one needs 
a rather large population of stars to find a sufficient number
of Mira variables to 
perform a statistical study on them. This means that a large area should 
be observed repeatedly for a long enough duration to cover several pulsational 
periods, usually a hard task to do.
However, the situation has been changed in the last decade by
the gravitational microlensing search projects:
Optical Gravitational Lensing Experiment
(OGLE; Udalski, Kubiak \& Szymanski, 1997), 
MAssive Compact Halo Objects project (MACHO; \citealt{Alcock-2000}),
Microlensing Observations in Astrophysics (MOA; \citealt{Bond-2001}),
and so on.
Researchers investigating variable stars are now enjoying
the wealth of the data provided by these projects. 

The region studied most extensively is apparently the Large Magellanic Cloud: 
the discovery of five parallel sequences 
in the $\log P$-$K$ diagram by \citet{Wood-2000} was followed 
by many papers (Noda et al. 2002, 2004; Kiss \& Bedding, 2003, 2004;
\citealt{Groenewegen-2004}; Ita et al. 2004a, 2004b;
\citealt{Fraser-2005}). 
Following the conclusion of \citet{Glass-1981} and \citet{Feast-1989} that 
the scatter around period-magnitude relation of Mira variables is small
for bolometric or near-infrared magnitudes,
papers focused on infrared properties 
of Mira variables and those of other kinds of red variables.  
However, the Galactic bulge region has not 
received full attention except for the Baade windows
(\citealt{Schultheis-2001};
 Glass and Schultheis 2002, 2003; Schultheis, Glass \& Cioni, 2004).
Wray, Eyer \& Paczy\'nski (2004) 
analyzed OGLE variables over larger area of the bulge,
but they only discussed
small-amplitude variables and not Mira variables.
Wo\'zniak, McCowan \& Vestrand (2004) 
also made a detailed analysis of microvariability 
of Mira variables in the OGLE data, but no effort was dedicated to 
searching for infrared counterparts, which is 
indispensable to study the period-magnitude relation.

In this paper, we collected Mira variables
in the OGLE data and obtained their counterparts in
the 2 Micron All-sky Survey 
(2MASS) point source catalogue \citep{Curti-2003} and
the {\it Midcourse Space Experiment (MSX)}
 point source catalogue \citep{Egan-2003}.
After we present data analyses to obtain a catalogue
in Section \ref{sec:Data},
we discuss
their photometric properties (Section \ref{sec:Discuss1})
and their space distribution in the bulge (Section \ref{sec:Discuss2}).

\section{DATA}
\label{sec:Data}

\subsection{Extraction of Mira Variables from OGLE-II Data}

\begin{table}
\begin{minipage}{86mm}
\begin{center}
\caption{The 49 OGLE-II bulge fields. $l$ and $b$, galactic coordinate;
$N_{\rm Mira}$, the number of OGLE-II Mira variables
obtained in this paper; $d_{\rm RA}$ and $d_{\rm Dec}$,
offset between the coordinates in OGLE and 2MASS; $A_K$,
the derived extinction value. $N_{\rm Mira}$ with the superscript $*$ indicates
a couple of objects have overlaps with other fields
 (see Table \ref{tab:duplicate} and text).\label{tab:fields}}
\begin{tabular}{rrrrrrrr}
\hline
Fld. & $l$ & $b$ & $N_{\rm Mira}$
& $d_{\rm RA}$ & $d_{\rm Dec}$ & $A_K$ \\
 & (deg) & (deg) & & (arcsec) & (arcsec) & (mag) \\
\hline
1   &     1.08 & $-$3.62 &   32$^*$ & $-$0.12 &    0.49 & 0.21 \\
2   &     2.23 & $-$3.46 &   29$^*$ &    0.30 &    0.07 & 0.18 \\
3   &     0.11 & $-$1.93 &  102$^*$ & $-$0.50 &    0.10 & 0.33 \\
4   &     0.43 & $-$2.01 &   83$^*$ & $-$0.15 &    0.15 & 0.29 \\
5   &  $-$0.23 & $-$1.33 &  120$^*$ & $-$0.04 &    0.27 & 0.60 \\
6   &  $-$0.25 & $-$5.70 &   12     & $-$0.35 &    0.56 & 0.13 \\
7   &  $-$0.14 & $-$5.91 &    9     & $-$0.50 &    0.66 & 0.13 \\
8   &    10.48 & $-$3.78 &    8$^*$ & $-$0.79 &    0.45 & 0.22 \\
9   &    10.59 & $-$3.98 &    9$^*$ & $-$0.73 &    0.43 & 0.21 \\
10  &     9.64 & $-$3.44 &   16     & $-$0.79 &    0.80 & 0.24 \\
11  &     9.74 & $-$3.64 &   14     & $-$0.80 &    1.34 & 0.24 \\
12  &     7.80 & $-$3.37 &   24     & $-$0.49 &    0.28 & 0.24 \\
13  &     7.91 & $-$3.58 &   17     & $-$0.46 &    0.42 & 0.21 \\
14  &     5.23 &    2.81 &   37     &    0.32 &    0.24 & 0.29 \\
15  &     5.38 &    2.63 &   31     &    0.30 &    0.21 & 0.32 \\
16  &     5.10 & $-$3.29 &   26     & $-$0.13 & $-$0.20 & 0.22 \\
17  &     5.28 & $-$3.45 &   31     & $-$0.17 & $-$0.42 & 0.21 \\
18  &     3.94 & $-$3.14 &   33$^*$ &    0.05 & $-$0.15 & 0.20 \\
19  &     4.08 & $-$3.35 &   34$^*$ & $-$0.02 & $-$0.16 & 0.21 \\
20  &     1.68 & $-$2.47 &   59$^*$ &    0.61 & $-$0.08 & 0.22 \\
21  &     1.80 & $-$2.66 &   53$^*$ &    0.55 & $-$0.03 & 0.22 \\
22  &  $-$0.26 & $-$2.95 &   60     & $-$0.47 &    0.46 & 0.31 \\
23  &  $-$0.50 & $-$3.36 &   49     & $-$0.46 &    0.55 & 0.30 \\
24  &  $-$2.44 & $-$3.36 &   40     &    0.74 &    0.17 & 0.32 \\
25  &  $-$2.32 & $-$3.56 &   44     &    0.52 &    0.35 & 0.28 \\
26  &  $-$4.90 & $-$3.37 &   30     & $-$0.68 & $-$0.23 & 0.23 \\
27  &  $-$4.92 & $-$3.65 &   23     & $-$0.69 & $-$0.30 & 0.21 \\
28  &  $-$6.76 & $-$4.42 &   12     & $-$1.33 & $-$0.32 & 0.20 \\
29  &  $-$6.64 & $-$4.62 &   11     & $-$1.36 & $-$0.11 & 0.19 \\
30  &     1.94 & $-$2.84 &   46$^*$ &    0.47 & $-$0.02 & 0.22 \\
31  &     2.23 & $-$2.94 &   52$^*$ &    0.46 &    0.03 & 0.20 \\
32  &     2.34 & $-$3.14 &   39$^*$ &    0.38 & $-$0.10 & 0.18 \\
33  &     2.35 & $-$3.66 &   31$^*$ &    0.31 &    0.11 & 0.20 \\
34  &     1.35 & $-$2.40 &   51$^*$ &    0.53 &    0.00 & 0.26 \\
35  &     3.05 & $-$3.00 &   28     &    0.30 & $-$0.18 & 0.21 \\
36  &     3.16 & $-$3.20 &   26     &    0.37 & $-$0.12 & 0.18 \\
37  &     0.00 & $-$1.74 &  104$^*$ &    0.04 &    0.30 & 0.40 \\
38  &     0.97 & $-$3.42 &   39$^*$ &    0.01 &    0.40 & 0.22 \\
39  &     0.53 & $-$2.21 &   94$^*$ & $-$0.03 &    0.24 & 0.30 \\
40  &  $-$2.99 & $-$3.14 &   55$^*$ & $-$0.53 &    0.10 & 0.36 \\
41  &  $-$2.78 & $-$3.27 &   36$^*$ &    0.84 &    0.09 & 0.33 \\
42  &     4.48 & $-$3.38 &   23     & $-$0.06 & $-$0.16 & 0.24 \\
43  &     0.37 &    2.95 &   70     &    0.48 &    0.26 & 0.44 \\
44  &  $-$0.43 & $-$1.19 &  140$^*$ & $-$0.12 &    0.28 & 0.77 \\
45  &     0.98 & $-$3.94 &   27     & $-$0.28 &    0.42 & 0.19 \\
46  &     1.09 & $-$4.14 &   28     & $-$0.10 &    0.50 & 0.19 \\
47  & $-$11.19 & $-$2.60 &   13     & $-$0.42 & $-$0.25 & 0.25 \\
48  & $-$11.07 & $-$2.78 &   11     & $-$0.52 & $-$0.19 & 0.24 \\
49  & $-$11.36 & $-$3.25 &    7     & $-$0.12 &    0.04 & 0.20 \\
\hline
\end{tabular}
\end{center}
\end{minipage}
\end{table}

The Optical Gravitational Lensing Experiment-II (OGLE-II)
repeatedly observed 49 fields between $l=-11\degr$
and $+11\degr$ from 1997 to 1999 \citep{Wozniak-2002}.
Galactic coordinates for the fields are given in Table \ref{tab:fields}.
Light curves (typically with 200--300 data points) for 221~801 objects 
in these fields are available on
the Internet.\footnote{OGLE homepage (http://www.astrouw.edu.pl/\%7Eogle/)}
After we excluded false data with error flags, we applied the phase 
dispersion minimization method \citep{Stellingwerf-1978}
to all the light curves.
We used the same software as \citet{Ita-2004a} did.
It gives us a period ($P$),
a regularity indicator of variation ($\theta$), 
and a maximum-to-minimum amplitude ($\Delta I$) for each variable.
Small $\theta$ indicates regular variation and a well-characterized period. 
We put the following criteria for Mira variables: 
\begin{eqnarray}
P &\geq& 100 ~{\rm days}, \label{eq:conP} \\
\theta &\leq& 0.6, \label{eq:conT} \\
\Delta I &\geq& 1, \label{eq:conA}
\end{eqnarray}
and obtained 11~207 candidates.
All the light curves were further examined by eye.
We found many fake detections that show very similar light curves
but are fainter than the corresponding real detection 
at neighbouring positions, and rejected them.
We also rejected some other candidates that showed double periodicity
which is often seen
among variables on the D sequence
(Mira variables are on the C sequence; \citealt{Wood-2000}).
Finally, we obtained 1968 Mira variables.
Table \ref{tab:Mira} shows the first 10 lines of
the resultant table. An entry with Fld$=f$ and ID$=N$ corresponds to
OGLE data named 'bul\_sc$f$\_$N$.dat4'. Their properties of variation
($P$, $\theta$, $\Delta I$, and mean magnitude $<I>$) are accompanied by
identification with 2MASS and {\it MSX} point source, which will be discussed
in the following sections.
The full version of the table will be electronically available 
in the online version of this journal.
The number of Mira variables for each field, $N_{\rm Mira}$,
is listed in Table \ref{tab:fields}.

\begin{table*}
\begin{minipage}{165mm}
\begin{center}
\caption{The first 10 lines in the result table of Mira variables.
This is a sample of the full version, which will be available in the online
version of this journal. In case of null identification with 2MASS or
{\it MSX} point source, 99.99 and 0.0 are put for magnitudes
and colours, respectively. Flag$=1$ indicates the
double detection in the neighbouring fields (otherwise 0;
see Table. \ref{tab:duplicate}).\label{tab:Mira}}
\begin{tabular}{rrccrrrrrrrrr}
\hline
Fld. & ID & 2MASS & MSX6C & $P$ & $\theta$ & $\Delta I$ & $<I>$ & $J-K_{\rm s}$ & $H-K_{\rm s}$ & $K_{\rm s}$ & [8] & Flag \\
& & & & days & & mag & mag & mag & mag & mag & mag & \\
\hline
 1 &   98 & 18022945-3024145 & ----------------- 
	& 244.3 & 0.26 & 5.29 & 13.43 & 1.50 & 0.50 &  7.56 & 99.99 &  0 \\
 1 &  103 & 18023199-3024487 & G000.5829-03.9157 
	& 349.6 & 0.10 & 2.23 & 11.72 & 1.75 & 0.69 &  7.12 &  5.11 &  0 \\
 1 &  177 & 18023020-3023485 & G000.5942-03.9016 
	& 332.3 & 0.36 & 6.21 & 14.18 & 1.92 & 0.70 &  6.90 &  4.37 &  0 \\
 1 &  193 & 18024239-3023403 & G000.6176-03.9391 
	& 336.5 & 0.10 & 2.02 & 12.55 & 1.91 & 0.72 &  7.21 &  5.38 &  0 \\
 1 &  235 & 18021264-3022350 & G000.5814-03.8367 
	& 264.7 & 0.10 & 1.93 & 11.62 & 1.60 & 0.60 &  7.58 &  5.92 &  0 \\
 1 &  268 & 18024758-3022355 & G000.6419-03.9480 
	& 180.4 & 0.28 & 1.66 & 11.14 & 1.40 & 0.50 &  7.81 &  6.23 &  0 \\
 1 &  558 & 18025358-3019229 & G000.7004-03.9399 
	& 230.0 & 0.08 & 2.83 & 13.68 & 2.04 & 0.84 &  8.08 &  5.85 &  0 \\
 1 &  602 & 18023988-3018221 & G000.6907-03.8889 
	& 315.9 & 0.05 & 3.44 & 12.65 & 1.62 & 0.57 &  6.87 &  5.46 &  0 \\
 1 &  878 & 18023810-3015202 & ----------------- 
	& 127.8 & 0.22 & 1.29 & 12.38 & 1.32 & 0.43 &  8.73 & 99.99 &  0 \\
 1 &  916 & 18021033-3014173 & G000.6978-03.7619 
	& 385.5 & 0.19 & 1.37 & 11.39 & 1.92 & 0.69 &  5.28 &  4.06 &  0 \\
\hline
\end{tabular}
\end{center}
\end{minipage}
\end{table*}

\subsection{Cross-identification with 2MASS\label{sec:2MASS}}
\begin{figure}
\begin{center}
\includegraphics[clip,width=0.99\hsize]{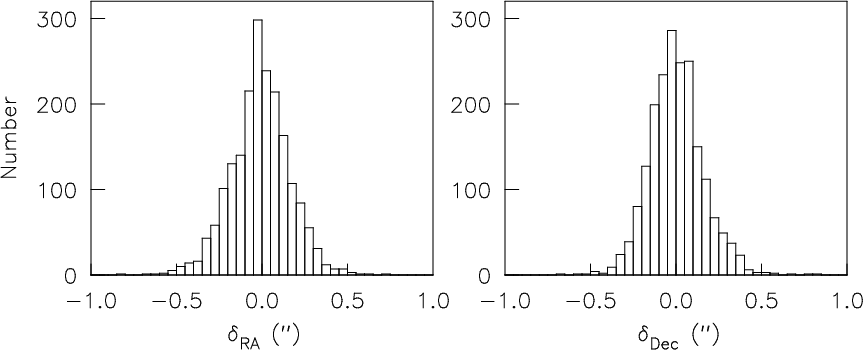}
\caption{Histogram of the positional differences between OGLE and 2MASS
coordinates for the cross-identified Mira variables.\label{fig:2MASSpos}}
\end{center}
\end{figure}

Near-infrared counterparts of the OGLE-II Mira variables were searched for 
in the 2MASS point source catalogue \citep{Curti-2003}.
In order to adjust the systematic differences in positions between the two 
data sets, we adopted the different offsets in right ascension and declination 
for different fields
($d_{\rm RA}$ and $d_{\rm Dec}$, given in Table \ref{tab:fields}).
With these offsets, we found 1960 matches within
a limit radius of 1 arcsec.
Fig. \ref{fig:2MASSpos} shows the differences in the right ascension and 
declination. It is easily seen that most of the OGLE-II Mira variables have 
their near-infrared counterparts within a radius of 0.5 arcsec. 

Light variability causes
main photometric uncertainties under discussion in this paper.
Although the amplitude of light variation decreases 
with the wavelength for Mira variables, some of them still have
an amplitude of up to 1 mag in the $K$-band (\citealt{Glass-1995}; 
Whitelock, Marang \& Feast, 2000). 
Therefore, mean magnitudes are
usually used to investigate the infrared properties of Mira 
variables. As the 2MASS catalogue gives us single-epoch
magnitudes, unfortunately, we should be aware of 
the possible error caused by the difference between the mean
magnitude and the single-epoch one. 
This error is large in comparison with the difference
between photometric systems.
According to \citet{Carpenter-2001}, for example, 
the difference is less than 0.1~mag
between $K_s$ in the 2MASS system and $K$
in the Las Campanas Observatory (LCO) system,
in which system \citet{Ita-2004a} derived the $\log P$-$K$ relation
(used in eq. \ref{eq:PLR}).
Therefore, we ignore the difference
between $K$ and $K_s$ magnitudes among various systems
in the following discussions.

\subsection{Cross-identification with {\it MSX}\label{sec:MSX}}

\begin{figure}
\begin{center}
\includegraphics[clip,width=0.99\hsize]{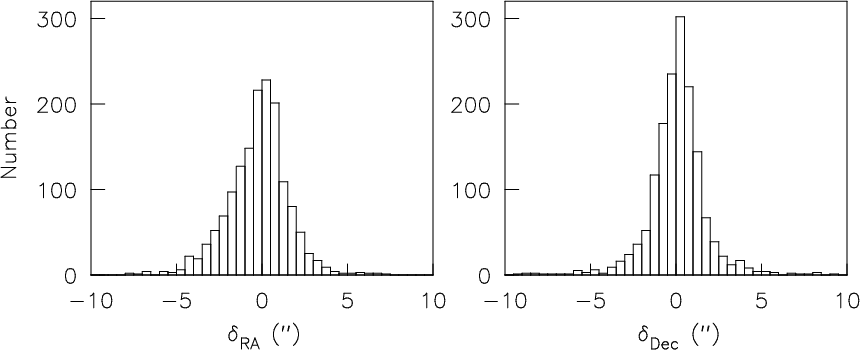}
\caption{Histogram of the positional differences between OGLE and {\it MSX}
coordinates for the cross-identified Mira variables.\label{fig:MSXpos}}
\end{center}
\end{figure}

Mid-infrared counterparts were searched for in the {\it MSX}
point source catalogue
\citep{Egan-2003}, leading to 1541 matches within
a limit radius of 10 arcsec. 
The positional differences are plotted in Fig. \ref{fig:MSXpos}.
Among the identifications with 2MASS and/or {\it MSX} point sources,
30 Mira variables were found to be doubly detected because they lie
in the overlapping regions between
the neighbouring OGLE-II fields, as listed in Table \ref{tab:duplicate}.
We put the flag 1 in the last column of Table \ref{tab:Mira}
if a variable has a counterpart in another field (otherwise 0).

\begin{table}
\begin{minipage}{86mm}
\begin{center}
\caption{Double detections in the overlapping regions between the neighbouring
fields.\label{tab:duplicate}}
\begin{tabular}{rrrrcc}
\hline
\multicolumn{2}{c}{Entry 1.} & \multicolumn{2}{c}{Entry 2.} & 2MASS & MSX6C \\
 Fld. & ID & Fld. & ID &   & \\
\hline
 1 & 3024 & 38 & 3446 & 18020050-2947556 & G001.0639-03.5153 \\
 2 & 3933 & 33 & 3377 & 18050058-2839528 & G002.3762-03.5326 \\
 3 &  878 & 37 &  857 & 17530423-3018592 & G359.6485-02.0981 \\
 3 & 2286 &  4 &  375 & 17540566-3009027 & G359.9034-02.2048 \\
 3 & 2695 & 37 & 2835 & 17530457-3005208 & G359.8450-01.9840 \\
 3 & 4124 &  4 & 2209 & 17540599-2956549 & G000.0783-02.1041 \\
 3 & 4235 &  4 & 2509 & 17540524-2955533 & ---------------- \\
 3 & 4522 & 37 & 4656 & 17530343-2952379 & G000.0255-01.8727 \\
 3 & 5929 & 37 & 5926 & 17530367-2944057 & G000.1497-01.8011 \\
 3 & 6181 &  4 & 4722 &----------------- & G000.2718-01.9909 \\
 3 & 7804 &  4 & 6265 & 17540688-2934463 & G000.3989-01.9203 \\
 4 &  360 & 39 &  315 & 17550808-3009531 & G000.0047-02.4062 \\
 5 & 84   & 44 &  805 & 17495238-3023328 & G359.2301-01.5432 \\
 5 & 367  & 44 & 1066 & 17495216-3020497 & ---------------- \\
 5 & 3681 & 44 & 4608 & 17495113-2954571 & G359.6381-01.2952 \\
 5 & 3916 & 44 & 5045 & 17495111-2952354 & G359.6717-01.2748 \\
 5 & 4498 & 44 & 5702 & 17495312-2948492 & G359.7288-01.2483 \\
 5 & 4968 & 44 & 6453 & 17495414-2944533 & G359.7876-01.2184 \\
 8 &  297 &  9 &  210 & 18233244-2206346 & ---------------- \\
18 & 1821 & 19 & 1540 & 18073300-2724048 & G003.7556-03.4096 \\
18 & 5324 & 19 & 4649 & 18073420-2649523 & G004.2579-03.1377 \\
20 &    1 & 34 & 1945 & 17584836-2920304 & ---------------- \\
21 & 2270 & 30 & 1627 & 18005408-2902532 & G001.5985-02.9364 \\
21 & 2895 & 30 & 2129 & 18005341-2858004 & G001.6689-02.8940 \\
21 & 5977 & 30 & 5077 & 18005369-2836083 & G001.9870-02.7149 \\
30 & 1821 & 31 &  159 & 18015154-2902159 & G001.7114-03.1135 \\
30 & 4589 & 31 & 1759 & 18015346-2840498 & G002.0268-02.9441 \\
31 &  793 & 32 &  877 & 18025457-2854244 & G001.9396-03.2480 \\
40 & 1674 & 41 &  957 & 17513443-3320272 & ---------------- \\
40 & 3769 & 41 & 3089 & 17513529-3251234 & G357.2979-03.1164 \\
\hline
\end{tabular}
\end{center}
\end{minipage}
\end{table}
The $A$-band (8.28 $\mu$m with the width of 3.36 $\mu$m) has
the highest sensitivity among the six photometric bands of the {\it MSX} survey.
The majority of identified objects have a quality flag 
$Q=4$ (Excellent) and all have flags better than $Q=2$ (Fair/Poor)
in the $A$-band,
while more than two thirds of objects have a flag of $Q=1$ (Limit) or
$Q=0$ (Not Detected) in other bands.
Therefore, we decided to use only the $A$-band flux 
in our discussions.
The $A$-band magnitudes (denoted as [8]) were obtained using
the zero magnitude flux of 58.49~Jy \citep{Egan-2003}.
Names and [8] magnitudes for identified sources are listed
in Table \ref{tab:Mira}.
The 427 objects without the {\it MSX} counterpart have 
relatively short periods ($P < 400$ days) and 
blue colours [ $(H-K)_0 < 1$ ],
i.e. fainter $K$ magnitudes according to the period-$K$ relation and
little excesses in mid infrared. 
They are naturally expected to be faint in the $A$-band.

\subsection{Interstellar extinction\label{sec:AK}}

\begin{figure}
\begin{center}
\includegraphics[clip,width=0.99\hsize]{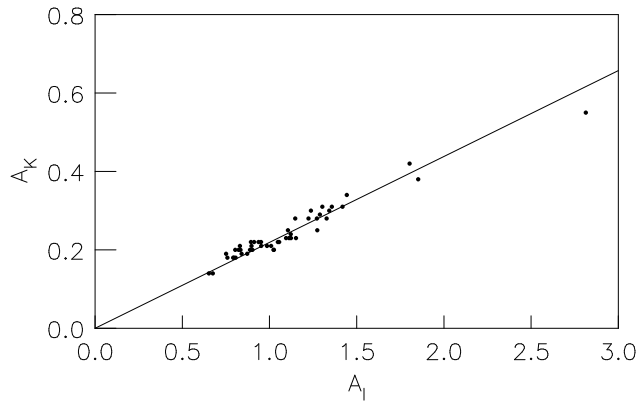}
\caption{Comparison of the extinction values between $A_I$ by \citet{Sumi-2004}
and $A_K$ by us.\label{fig:Excess}}
\end{center}
\end{figure}
The OGLE-II fields were chosen in the relatively low extinction regions in 
the Galactic bulge. Still they are located at low galactic latitudes where 
the interstellar extinction is generally strong.
In order to estimate the extinction for each OGLE-II field,
we measured how much the red giant branch shifts along a reddening vector
in the colour-magnitude diagram. Comparing the position of the branch
for each field with that for the Baade window NGC~6522 field,
the positional difference was interpreted as being the result of
the extinction difference between the two fields.
The extinction value for the NGC~6522 field itself was taken as $E_{B-V}=0.5$ 
from \citet{Schultheis-2004}.
The procedures were adopted on both the $J-K$ versus $K$ diagram
and the $H-K$ versus $K$ diagram.
Slopes of the reddening vectors were taken from \citet{Rieke-1985}:
\begin{eqnarray}
A_K / E(J-K) &=& 0.659,\\
A_K / E(H-K) &=& 1.78.
\end{eqnarray}
The extinction values obtained from both diagrams agree very well
within $\pm 0.02$ mag for every field.
The results are given in Table \ref{tab:fields}.
As shown in Fig. \ref{fig:Excess}, our results are consistent
with the $A_I$ values for OGLE-II fields obtained by \citet{Sumi-2004}.
Although a metallicity gradient within the bulge can cause shifts
on our estimates due to the dependence of the colour of giant branch,
such a gradient was not detected and is considered to be small
(\citealt{Ibata-1995};
Frogel, Tiede, \& Kuchinski, 1999; 
\citealt{Ramirez-2000}).
A difference of 0.2 dex in [Fe/H] results in a shift of 0.03 mag in $A_K$.
No reddening correction was applied to mid-infrared data
since they are small and uncertain.

\section{DISCUSSION: Photometric properties}
\label{sec:Discuss1}

\subsection{Completeness of the sample}
\begin{figure}
\begin{center}
\includegraphics[clip,width=0.99\hsize]{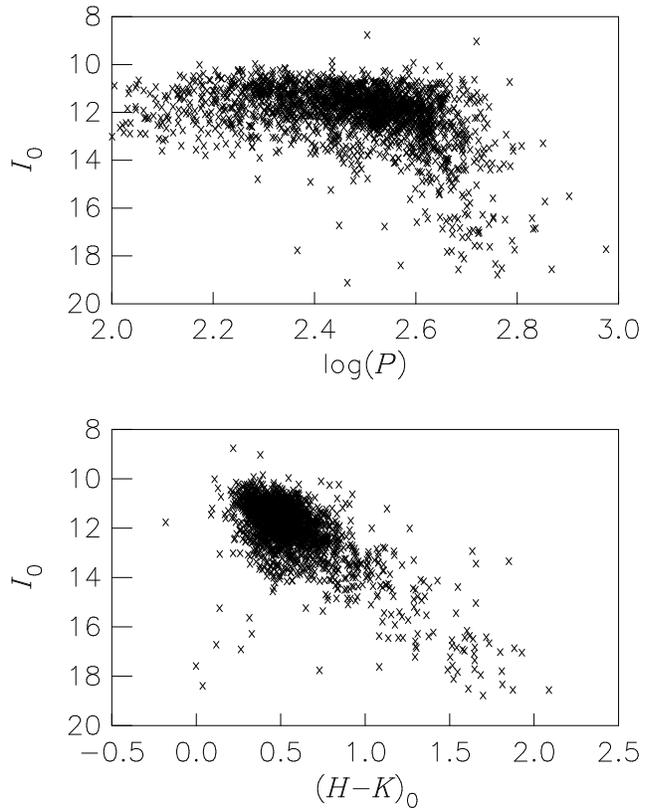}
\caption{$I_0$ magnitude plotted against $\log P$ (top)
and $(H-K)_0$ (bottom).\label{fig:Ifig}}
\end{center}
\end{figure}

\begin{figure}
\begin{center}
\includegraphics[clip,width=0.99\hsize]{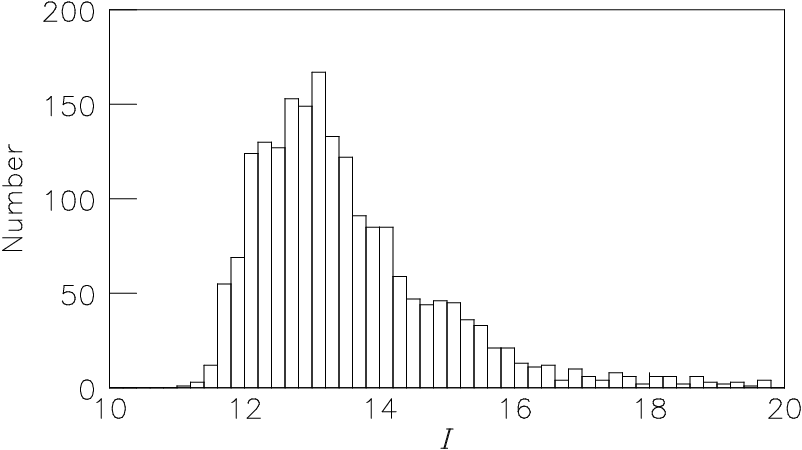}
\caption{Histogram of mean $I$-band magnitudes for the OGLE-II Mira variables.
\label{fig:MagHist}}
\end{center}
\end{figure}

The saturation limit of OGLE-II observation is about $I=11$.
As a result, the extinction-corrected 
$I_0$ magnitudes are limited to $I_0>10$ in Fig. \ref{fig:Ifig}. 
Some bright variables could be missing from our sample.
However, fewer than 10 per cent are brighter than 11~mag
in $I$-band magnitude distribution
of late-type M giants toward the Baade window obtained
by Blanco, McCarthy \& Blanco (1984). 
Their distribution shows a similar decrease toward brighter region ($I<12$)
as is shown by our histogram in Fig. \ref{fig:MagHist}.
Furthermore, kinematic studies showed
these bright Mira variables to have a small velocity dispersion,
and they are probably foreground objects in the outer bulge or disc
(\citealt{Feast-1987}; Sharples, Walker \& Cropper, 1990).

\begin{figure}
\begin{center}
\includegraphics[clip,width=0.99\hsize]{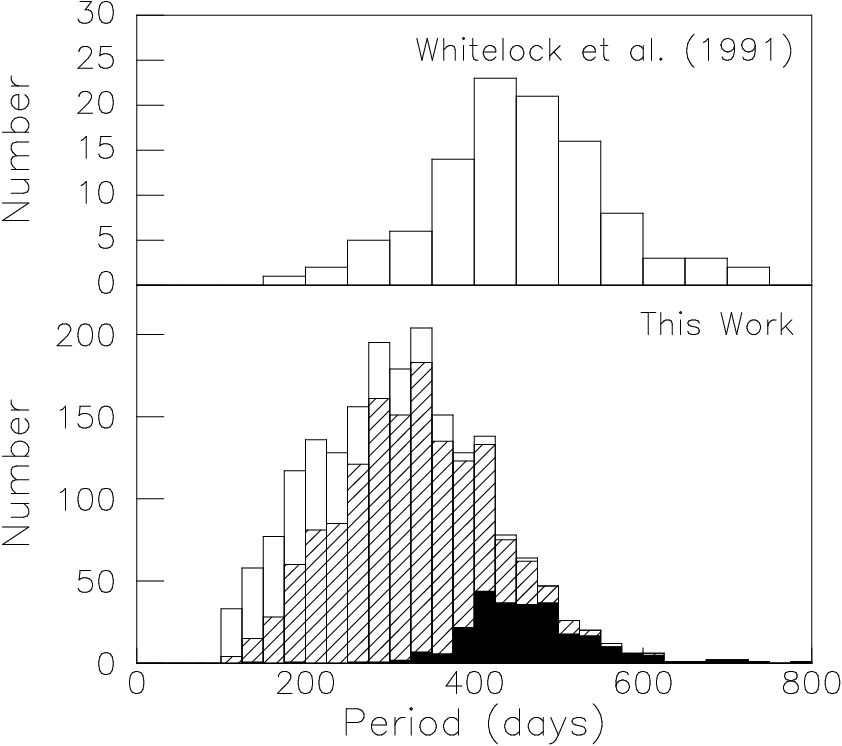}
\caption{
Histogram of periods for variables found in this work
and in the $7\degr <|b|<8\degr $ region by \citet{Whitelock-1991}.
In the bottom diagram, variables with {\it MSX} counterparts are hatched,
and among them ones with [8] $> 1$ Jy are filled.
\label{fig:Phist}}
\end{center}
\end{figure}

On the other extreme, no variables fainter than $I_0=19$ were detected. 
As shown in Fig. \ref{fig:Ifig},
the fainter boundary for the short-period Mira variables is clearly
delineated, but the long-period Mira variables seem to extend to below the 
detection limit. While the reddest sources in our sample show $(H-K)_0 \sim$ 2, 
some OH/IR stars near the Galactic Centre have as large $(H-K)_0$ as 3~mag 
(Wood, Habing \& McGregor, 1998). 
Such red objects have also been found among the {\it IRAS} sources 
in the Galactic bulge between $7\degr <|b|< 8\degr$
(Whitelock, Feast \& Catchpole, 1991). 
These extremely red objects are undergoing heavy mass loss which leads to 
large $(H-K)_0$. In such cases, $I$-band flux should be reduced to
below the detection limit. 
Fig. \ref{fig:Phist} compares the period distribution of our sample with 
that for the {\it IRAS} sample by \citet{Whitelock-1991}.
While their sample lacks short-period, and probably blue, Mira variables,
it contains more significant component at $P >$ 700~days than our sample.
Probably, some long-period red variables are missing from our sample.
In this regard, it is interesting that the OGLE-II objects with [8] $>1$ Jy 
(filled portion in Fig. \ref{fig:Phist}) have a more similar period
distribution to the {\it IRAS} sources. 

\subsection{Mass-losing O-rich star or C-rich star ?}

\begin{figure}
\begin{center}
\includegraphics[clip,width=0.99\hsize]{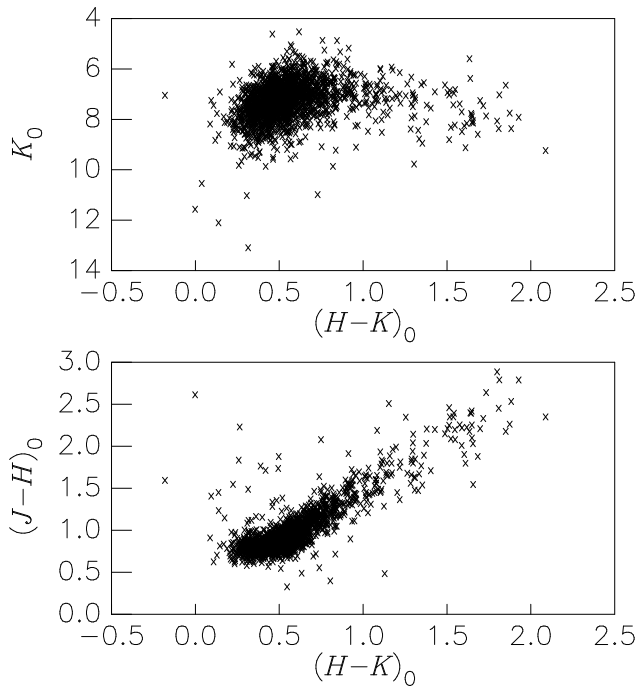}
\caption{Colour-magnitude diagram (top)
and colour-colour diagram (bottom)
for the Mira variables.\label{fig:2d}}
\end{center}
\end{figure}

\begin{figure}
\begin{center}
\includegraphics[clip,width=0.99\hsize]{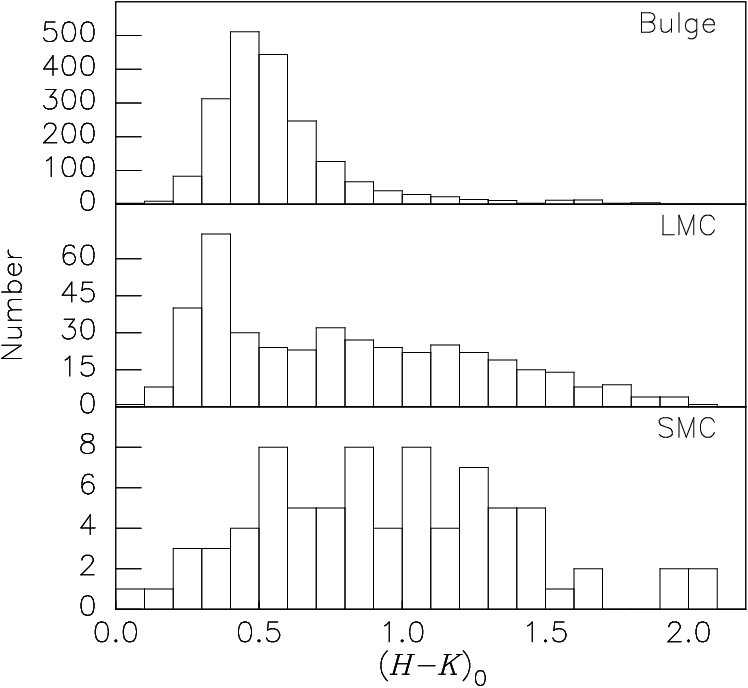}
\caption{Histogram of the colours $(H-K)_0$ of Mira variables
for the bulge (this work) and the both Magellanic Clouds \citep{Ita-2004b}.
\label{fig:HKHist}}
\end{center}
\end{figure}

\begin{figure}
\begin{center}
\includegraphics[clip,width=0.99\hsize]{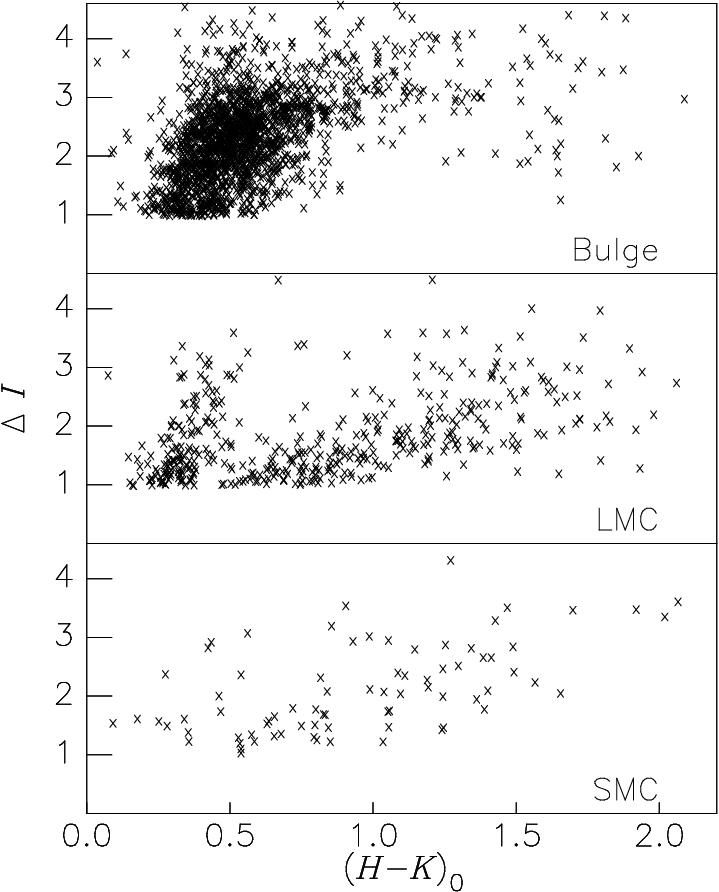}
\caption{Distributions on the colour $(H-K)_0$ versus the amplitude $\Delta I$
diagrams for the three regions.\label{fig:ColAmp}}
\end{center}
\end{figure}

\citet{Schultheis-2004} compared the
red variables in three regions,
namely the NGC~6522 field, the Large Magellanic Cloud (LMC),
and the Small Magellanic Cloud (SMC).
In the NGC~6522 field, they noted the absence of
variables with $(H-K)_0>1$, which are rather common
in both Magellanic Clouds (MCs).
They assigned those red objects to carbon-rich (C-rich)
stars which are absent in the Galactic bulge. 
Fig. \ref{fig:2d} shows the colour-magnitude and the colour-colour diagrams 
of the Mira variables in the OGLE-II fields. In contrast to the 
NGC~6522 field, our sample shows a red tail extending to $(H-K)_0=2$. 
However, Fig. \ref{fig:HKHist} shows the fraction of such red variables
is so small that the difference is not significant.
Fig. \ref{fig:HKHist} also presents the colour distribution of Mira variables
in both MCs, the data for which were collected from \citet{Ita-2004b}
using the same criteria, i.e. (\ref{eq:conP})--(\ref{eq:conA}).
As \citet{Schultheis-2004} mentioned, the distribution is clearly different
among three galaxies. This is caused by
the different combinations of oxygen-rich (O-rich) stars and
C-rich stars, as is discussed in the following.

Fig. \ref{fig:ColAmp} shows the relationship between the colour $(H-K)_0$
and the amplitude $\Delta I$.
\citet{Ita-2004b} argued that the dichotomy for the LMC is caused by
the difference between O-rich stars (bluer) and C-rich ones (redder).
A distinction between O- and 
C-rich stars is not apparent in Fig. \ref{fig:HKHist}, but is clearly seen 
in Fig. \ref{fig:ColAmp} as two separate branches. 
While the distribution of O-rich Mira variables forms a rather
sharp ridge at $0.3 < (H-K)_0 < 0.4$ for the LMC,
the ridge is wider and stretches to a redder colour for the bulge.
This can be explained by
the fact that the giant branch star with the higher metal 
content has the redder colour as shown for Galactic globular clusters
(Frogel, Cohen \& Persson, 1983; 
\citealt{Ferraro-2000}). 
On the other hand, the peak disappears for the SMC.
This is probably because the lower metal content for the SMC
reduces the fraction of O-rich stars. 

\begin{figure}
\begin{center}
\includegraphics[clip,width=0.99\hsize]{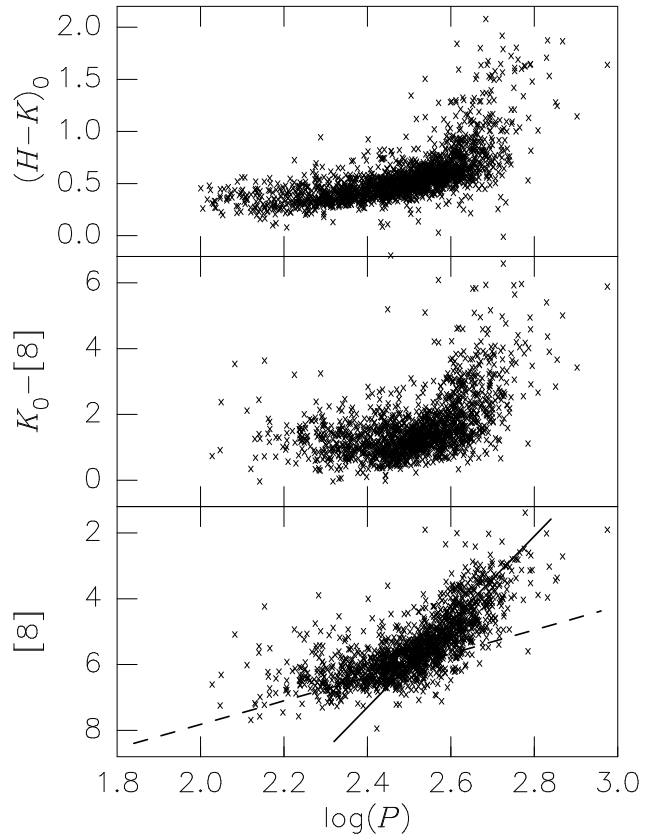}
\caption{
Relationships between period and colour
(top, $(H-K)_0$; middle, $K_0-[8]$)
and period-[8] magnitude relation.
Slopes of a solid line and a dashed line
are $-13$ and $-3.59$ respectively.
\label{fig:PLCPLA}}
\end{center}
\end{figure}

The dichotomy in Fig. \ref{fig:ColAmp} for the LMC disappears for the bulge.
The O-rich branch dominates, and the redder part looks extended from the branch.
This is in accordance with the fact that so far no bright
carbon star at the tip of the AGB has been found within the Galactic bulge
\citep{Blanco-1989}.
The top of Fig. \ref{fig:PLCPLA} shows that Mira variables in the redder
part have longer periods.
The middle panel and the bottom one plot
the colour $K_0-[8]$ and the mid-infrared magnitude [8], respectively,
against $\log P$.
These panels show Mira variables with periods longer than $\log P=2.55$
have [8]-band excesses indicating thick dust shells produced by
mass-loss phenomena. It is interesting that the $\log P$-[8] relation
clearly has a turn-off in contrast to the $\log P$-[12] relation
by \citet{Whitelock-1991} with only long period variables ($\log P>2.4$).
The slope turns up to a larger one ($-13$) from a value similar to that of
the $\log P$-$K$ relation ($-3.59$).

\begin{figure}
\begin{center}
\includegraphics[clip,width=0.99\hsize]{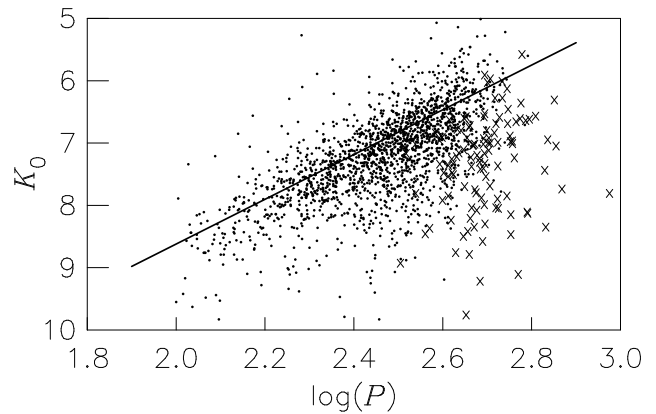}
\caption{The period-magnitude relation for the bulge Mira variables.
Crosses for Mira variables with $(H-K)_0\geq1$,
and filled circles for the others.\label{fig:PLR_K}}
\end{center}
\end{figure}

\begin{figure}
\begin{center}
\includegraphics[clip,width=0.99\hsize]{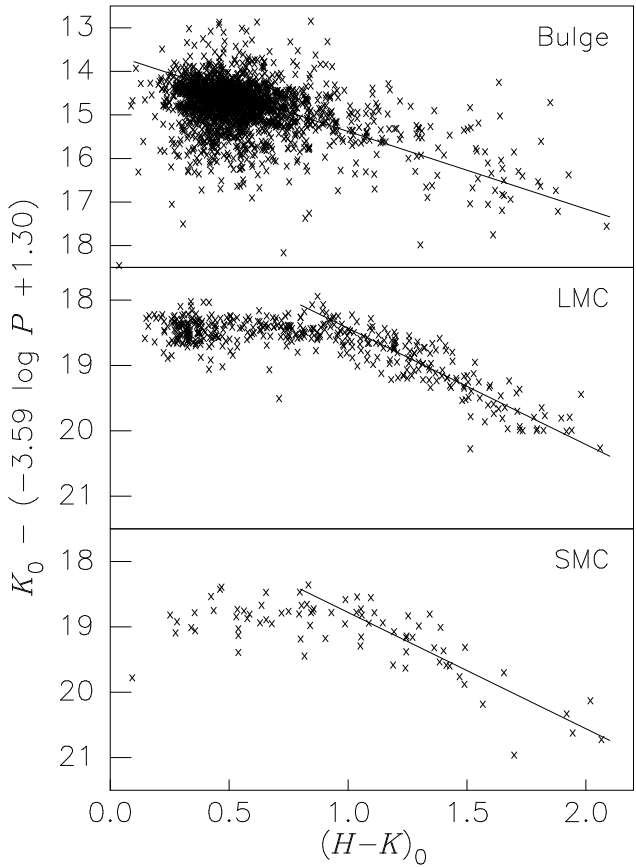}
\caption{The deviation from $\log P$-$M_K$ relation plotted against
the colour $(H-K)_0$ for the three regions. Solid lines are fitted
for the red part with $(H-K)_0\geq 1$.\label{fig:ColPLR}}
\end{center}
\end{figure}

\subsection{The $\log P$-$K$ relation\label{sec:PLR}}

Fig. \ref{fig:PLR_K} shows the $\log P$-$K_0$ diagram for
the OGLE-II Mira variables. The solid line is a $\log P$-$K$ relation 
of the LMC Mira variables (\citealt{Ita-2004a}) 
after the correction of difference in the distance modulus.
Crosses representing Mira variables with
$(H-K)_0 \geq 1$ occupy a fainter region because 
of the circumstellar extinction in the $K$-band.
Fig. \ref{fig:ColPLR} shows the difference between $K_0$ and
the $\log P$-absolute $K$ relation:
\begin{equation}
M_K = -3.59\log P +1.30,
\label{eq:PLR}
\end{equation}
for each variable, against $(H-K)_0$.
For the both MCs, it is clearly shown that
Mira variables with $(H-K)_0 \geq 1$ deviate from the constant values.
The circumstellar extinction is responsible for this $K$-band darkening
because the reddened variables have longer periods
as those in the bulge do (see Fig. \ref{fig:PLCPLA}).

The distribution for the bulge seems different
from those for the MCs and the deviations seem to occur even
for bluer Mira variables.
The distribution for the bulge
has a large spread compared with the MCs. This is caused by the depth
of the bulge along the line of sight. As stars in the bulge
have a rather wide range of metallicities (\citealt{Zoccali-2003},
and references therein), there could also be an effect from
some kinds of metallicity dependence. Although \citet{Ita-2004a}
claimed they found the metallicity dependence of the
zero point between both MCs, it is still a subject of controversy
(see \citealt{Feast-2004}, for a review). 

Solid lines in Fig. \ref{fig:ColPLR} are least-square fitted lines for
Mira variables with $(H-K)_0\geq 1$ after one 2-$\sigma$ clipping.
It is interesting that their slopes are similar to 
the reddening coefficient $A_K/E(H-K)$
($1.67\pm 0.20$ for the bulge; $1.67\pm 0.07$ for the LMC;
$1.73\pm 0.17$ for the SMC). This indicates circumstellar matters around
Mira variables have optical characteristics similar to those of
interstellar matters in the near infrared range ($JHK$).

\section{DISCUSSION: The structure of the bulge}
\label{sec:Discuss2}

\subsection{Surface number density }
\begin{figure}
\begin{center}
\includegraphics[clip,width=0.99\hsize]{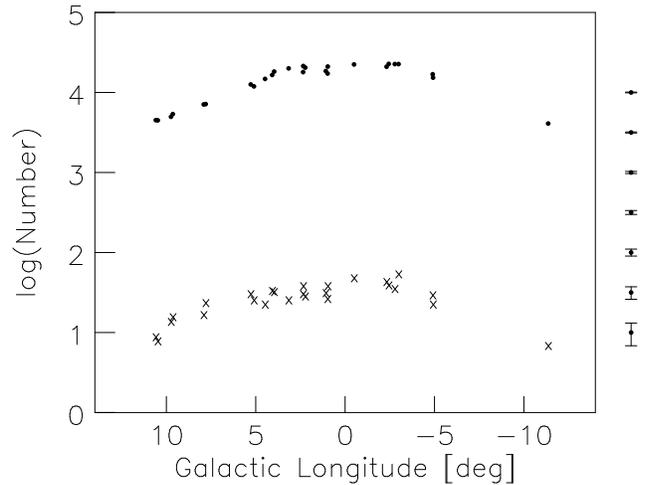}
\caption{The galactic longitude versus the number of Mira variables (cross)
and that of red clump stars (filled circle) for each OGLE-II field
with $-4\degr \leq b \leq -3\degr$\label{fig:Fld}}
\end{center}
\end{figure}

In Fig. \ref{fig:Fld}, the number of Mira variables for each OGLE-II field with
$-4\degr < b <-3\degr$ is plotted
against the galactic longitude (cross).
The amount of Poisson noise along the $y$-axis is illustrated 
on the right-hand side of the figure. 
The distribution is rather flat between $-3\degr < l <3\degr$ 
and drops to the half maximum at around $|l|=5\degr$.
We also confirmed this feature in the distribution of red clump stars
(filled circles in Fig. \ref{fig:Fld})
using the result obtained by \citet{Sumi-2004}.
This is because the surface brightness of the bulge is a so-called 'boxy'-type
and the brightness does not vary along $b=-3\degr$ between the region
(see, for example, fig. 1 by \citealt{Dwek-1995}).

\begin{figure}
\begin{center}
\includegraphics[clip,width=0.99\hsize]{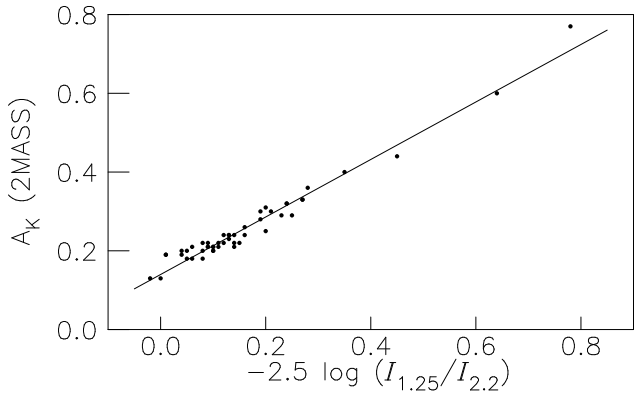}
\caption{The relationship between
{\it COBE}/DIRBE colours and the extinction values
obtained from 2MASS data (see Section \ref{sec:AK}).\label{fig:AkEJK}} 
\end{center}
\end{figure}

We compared the surface number density with the surface brightness 
obtained by the Diffuse Infrared Background Experiment (DIRBE)
onboard the {\it Cosmic Background Explorer} satellite {\it (COBE)}.
We used the Zodi-subtracted mission average map 
which is available
on the Web.\footnote{http://lambda.gsfc.nasa.gov/product/cobe/}
In order to extract the surface brightness of the bulge
(excluding the disc), we basically followed the method of \citet{Weiland-1994}.
First, we need to know the interstellar extinction.
Between the $A_K$ obtained in Section \ref{sec:AK} and 
the {\it COBE} colours at the OGLE-II fields, we found the relation
\begin{equation}
A_K = 0.73 \times [ -2.5 \log (I_{1.25}/I_{2.2}) + 0.14],
\label{eq:Extinction}
\end{equation}
as shown in Fig. \ref{fig:AkEJK}.
From this equation, we obtained $(J-K)_0=0.8$
as the intrinsic colour of the bulge population.
It is a reasonable colour for old and metal-rich stellar population.
For example, a recent population synthesis model by \citet{Mouhcine-2003}
predicts a stellar population with $Z=0.008$ (or [M/H]$=-0.4$)
would have such a colour over a wide range of ages between 16~Gyr and 2~Gyr.
We regard the above equation as being applicable to
areas other than those observed by the OGLE-II survey,
and used this relation to
estimate the interstellar extinction.

\begin{figure}
\begin{center}
\includegraphics[clip,width=0.99\hsize]{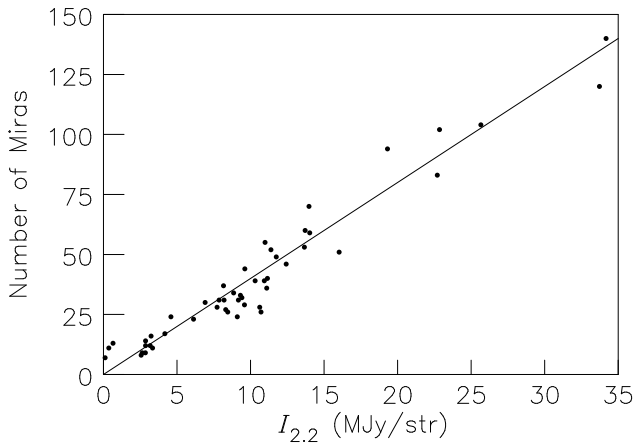}
\caption{
The relationship between the surface brightness of the bulge
obtained from {\it COBE}/DIRBE
and the number of Mira variables for each OGLE-II field.
\label{fig:COBE}}
\end{center}
\end{figure}

After the correction of the interstellar extinction,
we subtracted the disc contribution. The disc is described in
an exponential form:
\begin{equation}
I(l,b)_{\rm disc} = I(0,b)_{\rm disc} {\rm e}^{-|l|/l _0(b)},
\end{equation}
where $l_0(b)$ is the scale-hight for each galactic latitude, fitting to 
the observed surface brightness between $10\degr < |l| < 45\degr$.
With the obtained disc components subtracted,
the surface brightness of the bulge component for each field is obtained by 
averaging the values within $0.5\degr$ from
the central coordinate of the field.
Fig. \ref{fig:COBE} shows the relationship between
the thus-obtained surface brightness and the number of Mira variables
for each field. 

The specific frequency of Mira variables per unit apparent brightness
is 0.059 Jy$^{-1}$ from the slope of the figure.
Assuming a distance of 8~kpc for the bulge, the frequency per unit
$K$-band absolute brightness is 0.094 MJy$^{-1}$.
For metal-rich globular clusters, \citet{Frogel-1998} determined
the specific frequency of long-period variables to
be 0.11 MJy$^{-1}$ (see their eq. 2).
Considering that they didn't include variables with $P< 190~{\rm d}$,
whereas they account for about 10 per cent of our sample,
the frequency for the bulge is 25 per cent smaller than
that for the globular clusters. This can be because
the high metal content in the bulge enhances the mass loss and abbreviates
the lifetimes of Mira variables
as \citet{Frogel-1998} suggested.
The total number of Mira variables were estimated to be $6\times 10^5$
by integrating the surface brightness distribution
over the $|l|<10\degr$ and $|b|<10\degr$ region.

\subsection{Radial distribution}

The period-magnitude relation of Mira variables has been widely applied as
a distance indicator.
A recent example is \citet{Rejkuba-2004} where the relation is used to obtain 
the distance to the galaxy NGC~5128 (Cen~A).
Furthermore, Lah, Kiss \& Bedding (2005) 
explored the three-dimensional structures
of the MCs.
We expect that the structure of the bulge can also be traced by Mira variables.
It is considered to be shaped like a bar inclined from
the Sun-Galactic Centre line of sight by 15-40$\degr$
after many works with various methods
(distribution of luminous late-type stars, \citealt{Nakada-1991};
gas dynamics, \citealt{Binney-1991}; \citealt{Nakai-1992}; 
surface brightness distribution, \citealt{Blitz-1991}; \citealt{Dwek-1995};
distribution and kinematics of SiO masers, \citealt{Deguchi-2002};
peak of luminosity function due to red clump stars,
\citealt{Stanek-1994}; \citealt{Nishiyama-2005}).
Using 104 Mira variables, \citet{Whitelock-1992} also showed
the asymmetry of the bulge.

\begin{figure}
\begin{center}
\includegraphics[clip,width=0.99\hsize]{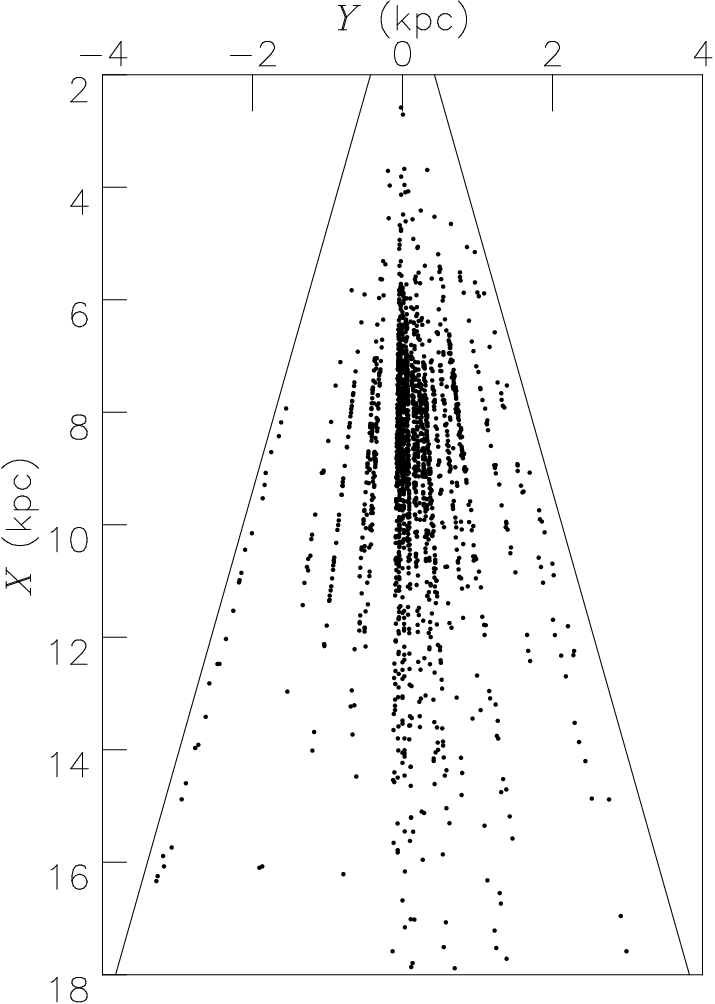}
\caption{Distribution of Mira variables projected on to the galactic plane.
The Sun locates at (0,0) and the Galactic Centre at around ($X=8,Y=0$).
Solid lines correspond to $l=\pm 12 \degr$.
\label{fig:Bar}} 
\end{center}
\end{figure}

In Fig. \ref{fig:Bar}, the distribution of OGLE-II Mira variables are 
projected on the galactic plane. 
Distances to individual variables 
are obtained by adopting the relation (\ref{eq:PLR}).
As the very red objects tend to deviate from this relation, 
we used only stars with $(H-K)_0\leq 1$
to avoid the circumstellar extinction.
Unfortunately, we had to use single-epoch magnitudes from the 2MASS catalogue,
instead of average magnitudes which are usually used for the distance
estimation. The uncertainty in our estimate from single-epoch magnitudes
causes the distance uncertainty to rise by up to $\pm 2$~kpc
in the extreme case.
Still we notice that there are far less stars at nearer region
to the Sun on the negative longitude side than on the positive side.
This asymmetry should reflect the elongated shape of the bulge.
It is expected that more dense surveys,
i.e. toward a number of galactic longitudes,
will reveal the detailed structure of the bulge.
It is also interesting to investigate the kinematics of the Mira variables
and to combine them with the space distribution.

\section{SUMMARY}
We collected 1968 Mira variables in the bulge from OGLE-II data base.
1960 objects were cross-identified in 2MASS point source catalogue,
and 1541 objects in {\it MSX} point source catalogue.
Interstellar extinctions for the near-infrared magnitudes of 2MASS were
corrected by our estimates of the extinction values $A_K$, which are
consistent with $A_I$ by \cite{Sumi-2004}. They are also consistent
with those obtained with the surface brightness colours from
{\it COBE}/DIRBE data.
Based on the catalogue of Mira variables,
some photometric properties were discussed.
Our results include:
(1) the separation between O-rich Mira variables and C-rich ones
in the colour-amplitude diagram,
(2) the preference for periods longer than $350~{\rm d}$ of mass-losing stars
demonstrated in the period-colour diagram and the period-[8] diagram,
and (3) that those mass-losing stars are affected by their own circumstellar
matter that have similar reddening vectors
to that of interstellar extinction.

We also discussed the structure of the bulge.
The number of Mira variables is proportional to
the {\it COBE}/DIRBE surface brightness, for which interstellar extinction 
and disc contribution were corrected.
Using the coefficient of proportionality, we estimated
the total number of Mira variables in the bulge is about $6\times 10^5$.
Our analysis provided a panoramic view of the bulge
although it is a preliminary result due to the lack of repeated observations
in the infrared.
However, we clearly detected the asymmetry between
the positive galactic-longitude side and the negative side,
which is commonly considered as evidence
of the bar-like shape of the bulge.

\section*{ACKNOWLEDGMENTS}
We are grateful to the OGLE project team who released
the large amount of data we used.
This publication also makes use of data products from
the Two Micron All-sky Survey,
which is a joint project between the University of Massachusetts and
the Infrared Processing and Analysis Center/California Institute of Technology,
funded by the National Aeronautics and Space Administration and
the National Science Foundation.
We further used data products from the Midcourse Space Experiment.
Processing of the data was funded by the Ballistic 
Missile Defense Organization with additional support from NASA
Office of Space Science.  This research has also made use of the
NASA/IPAC Infrared Science Archive, which is operated by the
Jet Propulsion Laboratory, California Institute of Technology,
under contract with the National Aeronautics and Space
Administration.
We acknowledge the use of the Legacy Archive for Microwave Background Data
Analysis (LAMBDA) for the {\it COBE} data. Support for LAMBDA is provided by
the NASA Office of Space Science.
One of the authors (MN) is financially supported by
the Japan Society for the Promotion of Science (JSPS) for Young Scientists.

\section*{Note added in proof}
Croenewegen \& Blommaert (2005) have also made an analysis of Mira variables
in the OGLE-II data, and independently obtained some similar results to ours.
In particular, they present estimations of the structure using simulations.

\section*{Supplementary material}
The following supplementary material is available online.

{\bf Table 2.} The result table of Mira variables. In case of null
identification with 2MASS or the MSX point source, 99.99 and 9.9 are put
for magnitudes and colours, respectively. Flag$=$ 1 indicates double
detection in the neighbouring fields (otherwise 0; see Table 3).

%
\label{lastpage}


\begin{thebibliography}{99}
\bibitem[\protect\citeauthoryear{Alcock et al.}{2000}]{Alcock-2000}
Alcock C. et al., 2000, ApJ, 541, 734
\bibitem[\protect\citeauthoryear{Blanco et al.}{1984}]{Blanco-1984}
Blanco, V.~M., McCarthy, M.~F., Blanco, B.~M., 1984, AJ, 89, 636
\bibitem[\protect\citeauthoryear{Blanco \& Terndrup}{1989}]{Blanco-1989}
Blanco, V.~M., Terndrup, D.~M., 1989, AJ, 98, 843
\bibitem[\protect\citeauthoryear{Bond et al.}{2001}]{Bond-2001}
Bond I.~A. et al., 2001, MNRAS, 327, 868
\bibitem[\protect\citeauthoryear{Binney et al.}{1991}]{Binney-1991}
Binney, J., Gerhard, O.~E., Stark, A.~A., Bally, J., Uchida, K.~I.,
1991, MNRAS, 252, 210
\bibitem[\protect\citeauthoryear{Blitz \& Spergel}{1991}]{Blitz-1991}
Blitz, L., Spergel, D.~N., 1991, ApJ, 379, 631
\bibitem[\protect\citeauthoryear{Carpenter}{2001}]{Carpenter-2001}
Carpenter, J.~M., 2001, AJ, 121, 2851
\bibitem[\protect\citeauthoryear{Curti et al.}{2003}]{Curti-2003}
Curti, R.~M. et al., 2003, Explanatory Supplement to
the 2MASS All Sky Data Release (Pasadena: Caltech)
\bibitem[\protect\citeauthoryear{Deguchi et al.}{2002}]{Deguchi-2002}
Deguchi, S., T. Fujii, T., Nakashima, J., Wood, P.~R., 2002, PASJ, 54, 719
\bibitem[\protect\citeauthoryear{Dwek et al.}{1995}]{Dwek-1995}
Dwek, E. et al., 1995, ApJ, 445, 716
\bibitem[\protect\citeauthoryear{Egan et al.}{2003}]{Egan-2003}
Egan, M.~P., et~al., 2003, The Midcourse Space
  Experiment Point Source Catalog Version 2.3, Air Force Research Laboratory
  Technical Report (AFRL-VS-TR-2003-1589)
\bibitem[\protect\citeauthoryear{Feast \& Whitelock}{1987}]{Feast-1987}
Feast, M.~W., Whitelock, P.~A., 1987, in
Kwok, S., Pottasch, S.~R., eds, Late stages of stellar evolution,
Reidel, Dordrecht, p. 33,
\bibitem[\protect\citeauthoryear{Feast et al.}{1989}]{Feast-1989}
Feast, M.~W., Glass, I.~S., Whitelock, P.~A., Catchpole, R.~M.,
1989, MNRAS, 241, 375
\bibitem[\protect\citeauthoryear{Feast}{2004}]{Feast-2004}
Feast, M.~W., 2004, ASPC, 310, 304
\bibitem[\protect\citeauthoryear{Ferraro et al.}{2000}]{Ferraro-2000}
Ferraro, F.~R., Montegriffo, P., Origlia, L., Fusi Pecci, F., 2000,
AJ, 119, 1282
\bibitem[\protect\citeauthoryear{Fraser et al.}{2005}]{Fraser-2005}
Fraser O.~J., Hawley S.~L., Cook, K.~H., Keller, S.~C., 2005, AJ, 129, 768
\bibitem[\protect\citeauthoryear{Frogel et al.}{1983}]{Frogel-1983}
Frogel, J.~A., Cohen, J.~G., Persson, S.~E., 1983, ApJ, 275, 773
\bibitem[\protect\citeauthoryear{Frogel et al.}{1999}]{Frogel-1999}
Frogel, J.~A., Tiede, G.~P., Kuchinski, L.~E., 1999, AJ, 117, 2296
\bibitem[\protect\citeauthoryear{Frogel \& Whitelock}{1998}]{Frogel-1998}
Frogel, J.~A., Whitelock, P.~A., 1998, AJ, 116, 754
\bibitem[\protect\citeauthoryear{Glass \& Lloyd Evans}{1981}]{Glass-1981}
Glass, I.~S., Lloyd Evans, T., 1981, Nature, 291, 303
\bibitem[\protect\citeauthoryear{Glass et al.}{1995}]{Glass-1995}
Glass, I.~S., Whitelock, P.~A., Catchpole, R.~M., Feast, M.~W., 1995,
MNRAS, 273, 383
\bibitem[\protect\citeauthoryear{Glass \& Schultheis}{2002}]{Glass-2002}
Glass, I.~S., Schultheis, M., 2002, MNRAS, 337, 519
\bibitem[\protect\citeauthoryear{Glass \& Schultheis}{2003}]{Glass-2003}
Glass, I.~S., Schultheis, M., 2003, MNRAS, 345, 39
\bibitem[\protect\citeauthoryear{Groenewegen}{2004}]{Groenewegen-2004}
Groenewegen, M.~A.~T., 2004, A\&A, 425, 595
\bibitem[\protect\citeauthoryear{Ibata \& Gilmore}{1995}]{Ibata-1995}
Ibata, R.~A., Gilmore, G.~F., 1995, MNRAS, 275, 605
\bibitem[\protect\citeauthoryear{Ita et al.}{2004a}]{Ita-2004a}
Ita, Y. et al., 2004a, MNRAS, 347, 720
\bibitem[\protect\citeauthoryear{Ita et al.}{2004b}]{Ita-2004b}
Ita, Y. et al., 2004b, MNRAS, 353, 705
\bibitem[\protect\citeauthoryear{Kiss \& Bedding}{2003}]{Kiss-2003}
Kiss, L.~L., Bedding, T.~R. 2003, MNRAS, 343, L79
\bibitem[\protect\citeauthoryear{Kiss \& Bedding}{2004}]{Kiss-2004}
Kiss, L.~L., Bedding, T.~R. 2004, MNRAS, 347, L83
\bibitem[\protect\citeauthoryear{Lah et al.}{2005}]{Lah-2005}
Lah, P., Kiss, L.~L., Bedding, T.~R., 2005, MNRAS, 359, L42
\bibitem[\protect\citeauthoryear{Mouhcine \& Lan\c{c}on}{2003}]{Mouhcine-2003}
Mouhcine, M., Lan\c{c}on, A., 2003, 402, 425
\bibitem[\protect\citeauthoryear{Nakai}{1992}]{Nakai-1992}
Nakai, N., 1992, PASJ, 44, L27
\bibitem[\protect\citeauthoryear{Nakada et al.}{1991}]{Nakada-1991}
Nakada, Y., Deguchi, S., Hashimoto, O., Izumiura, H., Onaka, T., 
Sekiguchi, K., Yamamura, I., 1991, Nature, 353, 140
\bibitem[\protect\citeauthoryear{Nishiyama et al.}{2005}]{Nishiyama-2005}
Nishiyama, S. et al., 2005, ApJ, 621, L105
\bibitem[\protect\citeauthoryear{Noda et al.}{2002}]{Noda-2002}
Noda, S. et al., 2002, MNRAS, 330, 137
\bibitem[\protect\citeauthoryear{Noda et al.}{2004}]{Noda-2004}
Noda, S. et al., 2004, MNRAS, 348, 1120
\bibitem[\protect\citeauthoryear{Ramir\'ez et al.}{2000}]{Ramirez-2000}
Ramir\'ez, S.~V., Stephens, A.~W., Frogel, J.~A., DePoy, D.~L,
2000, AJ, 120, 833
\bibitem[\protect\citeauthoryear{Rejkuba}{2004}]{Rejkuba-2004}
Rejkuba, M., 2004, A\&A, 413, 903
\bibitem[\protect\citeauthoryear{Rieke \& Lebfsky}{1985}]{Rieke-1985}
Rieke, G.~H., Lebofsky, M.~J., 1985, ApJ, 288, 618
\bibitem[\protect\citeauthoryear{Schultheis \& Glass}{2001}]{Schultheis-2001}
Schultheis, M., Glass, I.~S., 2001, MNRAS, 327, 1193
\bibitem[\protect\citeauthoryear{Schultheis et al.}{2004}]{Schultheis-2004}
Schultheis, M., Glass, I.~S. Cioni, M.-R., 2004, A\&A, 427, 945
\bibitem[\protect\citeauthoryear{Sharples et al.}{1990}]{Sharples-1990}
Sharples, R., Walker, A., Cropper, M., 1990, MNRAS, 1990, 246, 54
\bibitem[\protect\citeauthoryear{Stanek et al.}{1994}]{Stanek-1994}
Stanek, K. Z., Mateo, M., Udalski, A., Szyma\'nski, M.,
Kalu\'zny, J., Kubiak, M., 1994, ApJ, 429, L73
\bibitem[\protect\citeauthoryear{Stellingwerf}{1978}]{Stellingwerf-1978}
Stellingwerf, B.~F., 1978, ApJ, 224, 953
\bibitem[\protect\citeauthoryear{Sumi}{2004}]{Sumi-2004}
Sumi, T., 2004, MNRAS, 349, 193
\bibitem[\protect\citeauthoryear{Udalski et al.}{1997}]{Udalski-1997}
Udalski A., Kubiak M., Szymanski M., 1997, AcA, 47, 319
\bibitem[\protect\citeauthoryear{Weiland et al.}{1994}]{Weiland-1994}
Weiland, J.~L. et al., 1994, ApJ, 425, L81
\bibitem[\protect\citeauthoryear{Whitelock et al.}{1991}]{Whitelock-1991}
Whitelock, P.~A., Feast, M.~W., Catchpole, R.~M., 1991, MNRAS, 248, 276
\bibitem[\protect\citeauthoryear{Whitelock \& Catchpole}{1992}]{Whitelock-1992}
Whitelock, P.~A., Catchpole, R.~M., 1992,
in Blitz, L. eds, The center, bulge, and disk of the Milky Way,
Dordrecht, Kluwer, p. 103
\bibitem[\protect\citeauthoryear{Whitelock et al.}{2000}]{Whitelock-2000}
Whitelock, P.~A., Marang, F., Feast, M.~W., 2000, MNRAS, 319, 728
\bibitem[\protect\citeauthoryear{Wood et al.}{1998}]{Wood-1998}
Wood P.~R., Habing, H.~J., McGregor, P.~J., 1998, A\&A, 336, 925
\bibitem[\protect\citeauthoryear{Wood}{2000}]{Wood-2000}
Wood, P.~R., 2000, PASA, 17, 18
\bibitem[\protect\citeauthoryear{Wo\'zniak et al.}{2002}]{Wozniak-2002}
Wo\'zniak, P.~R., Udalski, A., Szymanski, M., Kubiak, M., Pietrzynski, G.,
Soszynski, I., Zebrun, K., 2002, Acta Astron., 52, 129.
\bibitem[\protect\citeauthoryear{Wo\'zniak et al.}{2004}]{Wozniak-2004}
Wo\'zniak, P.~R., McGowan, K.~E., Vestrand, W.~T., 2004, ApJ, 610, 1038
\bibitem[\protect\citeauthoryear{Wray et al.}{2004}]{Wray-2004}
Wray, J.~J., Eyer, L., Paczy\'nski, B., 2004, MNRAS, 349, 1059
\bibitem[\protect\citeauthoryear{Zoccali et al.}{2003}]{Zoccali-2003}
Zoccali, M. et al., 2003, A\&A, 399, 931
\end{thebibliography}
\end{document}